\documentclass[prb,twocolumn,floats,aps]{revtex4}
\usepackage{graphicx,graphics,color,epsfig} % Include figure files
\usepackage{bm}
\usepackage{amssymb}
\usepackage{amsmath}
\usepackage{amsfonts}

\begin{document}

\title{Persistence of energy-dependent localization in the Anderson-Hubbard model
with increasing system size and doping}
\author{P. Daley}
\author{R. Wortis}
%\author{J. Perera}
\affiliation{Department of Physics \& Astronomy, Trent University,
1600 West Bank Dr., Peterborough ON, K9J 7B8, Canada}
\date{\today}

\begin{abstract}
Non-interacting systems with bounded disorder have been shown to exhibit sharp density of states peaks at the band edge which coincide with an energy range of abruptly suppressed localization. %\cite{Johri2012a,Johri2012b}
Recent work %\cite{Perera2015} 
has shown that these features also occur in the presence of on-site interactions in ensembles of two-site Anderson-Hubbard systems at half filling.
Here we demonstrate that this effect in interacting systems persists away from half filling, and moreover that energy regions with suppressed localization continue to appear in ensembles of larger systems despite a loss of sharp features in the density of states.
\end{abstract}
\maketitle

\section{Introduction}
\label{sec-intro}

Following the discovery of Anderson localization\cite{Anderson1958} and given the importance of the density of states (DOS) as a tool for characterizing electronic behavior, there was interest in identifying DOS features associated with localization.
However, it was shown that for all continuous distributions over a wide energy range\cite{Edwards1971} and for the case of Gaussian distributed disorder for all energies\cite{Wegner1981} no sharp features arise in the DOS.  
Moreover, localization is generally strongest near the band edges where the DOS is lowest.
Recent work by Johri and Bhatt\cite{Johri2012a,Johri2012b} on the (non-interacting) Anderson model with a bounded disorder distribution was surprising on both of these points.
First, they found sharp features in the DOS, outside the energy range addressed in Ref.\ [\onlinecite{Edwards1971}].
Second, they found that these DOS peaks marked the boundary of a region at the band edge of sharply reduced localization as measured by inverse participation ratio, a phenomenon they associated with Lifshitz states.  
They observed these effects in large systems and in greater than one dimension,
and by considering an ensemble of two-site systems they were able to provide a precise, analytic explanation.\cite{Johri2012b}

The question of how interactions effect localization has been around since Anderson's original paper\cite{Anderson1958}
but recent rapid progress on the phenomenon of many-body localization\cite{Basko2006, Nandkishore2015,Altman2015}
has brought new attention to the topic, motivating the question:  Do these DOS and inverse participation features seen by Johri and Bhatt persist in the presence of interactions?
It has been shown very recently\cite{Perera2015} that indeed, in an ensemble of two-site Anderson-Hubbard systems at half filling, DOS peaks like those in non-interacting systems do occur in interacting systems and moreover that they are associated with regions of suppressed localization as measured by the generalized inverse participation ratio (GIPR).
Of particular interest is the fact that these regions of reduced localization move toward the Fermi level, making them more experimentally accessible.

\begin{figure*}
(a)
\includegraphics[height=1.4 in]{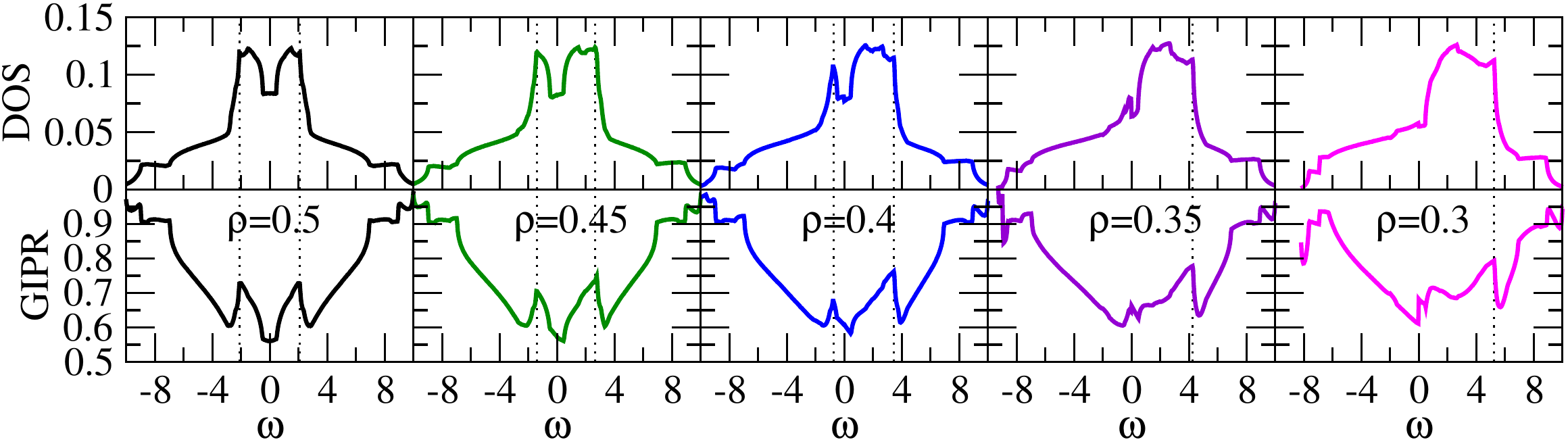} (b) 
\includegraphics[height=1.3 in]{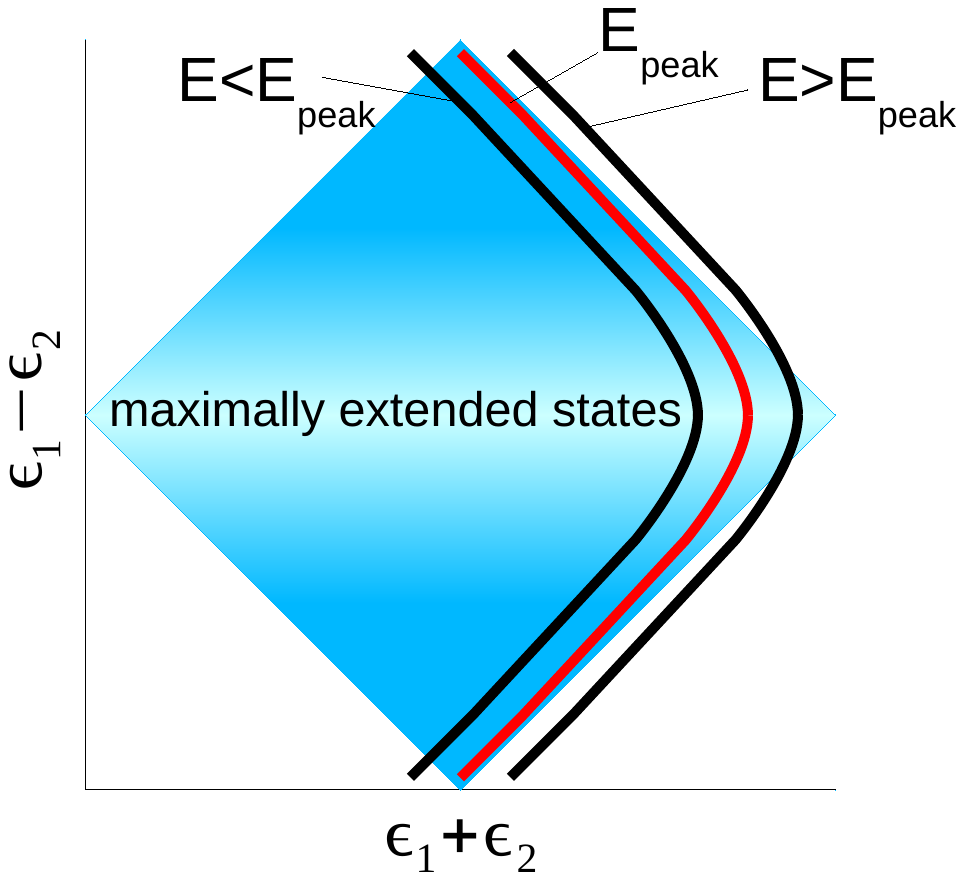}
\caption{\label{filling} (Color online)  (a) Variation with filling of the ensemble-average (a) DOS and (b) GIPR.  
Each curve is for an ensemble of 10 million two-site systems with disorder strength $W/t=12$ and interaction strength $U/t=8$.  
Chemical potentials $\mu/t=4$, 3.3, 2.6, 1.79 and 0.8 were used to obtain fillings $\rho=0.5$, 0.45, 0.4, 0.35 and 0.3 respectively.
Energy resolution is $0.04t$.
(b) Cartoon demonstrating the origin of the DOS peak and GIPR suppression in the simple case of no interactions.  See text for details. }
\end{figure*}

In this study we explore whether these features persist away from half filling and in ensembles of larger systems.
Indeed we find (Fig.\ \ref{filling} (a)) that sharp DOS peaks continue to correlate with dips in localization for ensembles of two-site systems at a range of fillings.
For larger system sizes (Fig.\ \ref{size}), the sharp DOS features are lost, but an energy range of reduced localization continues to appear.
Below we review the Anderson-Hubbard model and our calculations, and we present our results.

\section{The Anderson-Hubbard Model}
\label{sec-method}

A simple model for disordered systems is the non-interacting Anderson tight-binding model which combines nearest-neighbor hopping $t$ with site potentials chosen from a random distribution.
Among the simplest models for systems with both disorder and interactions is the Anderson-Hubbard model which adds an on-site Coulomb repulsion $U$.
\begin{eqnarray}
\mathcal{H}
&=& 
t \sum_{\langle i,j \rangle,\sigma} {\hat c}_{i\sigma}^{\dag} {\hat c}_{j\sigma}
+ \sum_i U {\hat n}_{i\uparrow} {\hat n}_{i\downarrow} 
+ \sum_{i,\sigma} (\epsilon_i - \mu) {\hat n}_{i\sigma}
\end{eqnarray}
${\hat c}_{i\sigma}^{\dag}$ is the creation operator for an electron with spin $\sigma$ at lattice site $i$,
${\hat n}_{i\sigma}={\hat c}_{i\sigma}^{\dag} {\hat c}_{i\sigma}$, and   
$\langle i,j \rangle$ refers to nearest neighbor pairs.
The site potentials $\epsilon_i$ are chosen from the distribution $P(\epsilon_i)=\Theta(W/2 - |\epsilon_i|)/W$ where $\Theta$ is the Heaviside function.  
$\mu$ is the chemical potential.

Here we consider ensembles of small systems with periodic boundary conditions. For two-site systems and four-site systems there is no distinction between one dimension and two dimensions.  For eight sites, we show results for a two-dimensional eight-site Betts lattice.\cite{Betts1996}
For each system in an ensemble, the site potentials are different but all are chosen from the same distribution.
Filling is controlled by the chemical potential.  Choosing $\mu=U/2$ results in half filling, but away from half filling no simple analytic expression is available.  The value of $\mu$ needed to obtain a desired filling for a given set of parameters is arrived at by iterative trials.  

\section{Calculations}
\label{sec-calc}

The eigenvalues $E_n$ and eigenstates $|\psi_n\rangle$ of each system are calculated by exact diagonalization using standard LAPACK routines.
% Patrick suggested mentioning binary number notation for Fock states and the automated construction of the Hamiltonian.  Bill felt this was sufficiently standard to not merit mention (and he wasn't aware of any quick technical name for it).
The many-body eigenstate with the lowest grand potential $\Omega_0=E_0-\mu N_0$ is the ground state, where $E_0$ and $N_0$ are the energy and particle number corresponding to this state.
All single-particle excitations from this ground state are considered in order to calculate the local retarded Green's function
\begin{eqnarray}
G_{ii\alpha\alpha}^R(\omega) &=& \sum_n \biggl\{
{|\langle \psi_n | {\hat c}_{i\alpha}^{\dag} |\psi_0\rangle|^2 \over \omega- (\Omega_n-\Omega_0) + i \eta} \nonumber \\
& & \hskip 0.3 in
+ {|\langle \psi_n | {\hat c}_{i\alpha} |\psi_0\rangle|^2 \over \omega + (\Omega_n-\Omega_0) + i \eta}
\biggr\}
\end{eqnarray}
Because the Hamiltonian conserves spin, $G_{ii\alpha\beta}^R(\omega)$ is zero for $\alpha \ne \beta$.
From this we calculate the local DOS at site $i$
\begin{eqnarray}
\rho_i(\omega) &=& - {1 \over \pi} {\rm Im} \ G_{ii}^R(\omega) \ \ {\rm where} \\
G_{ii}^R(\omega) &=& {1 \over 2} \left( G_{ii \uparrow \uparrow}^R(\omega) + G_{ii\downarrow \downarrow}^R(\omega) \right) 
\end{eqnarray}
The ensemble-average DOS is simply the average of $\rho_i(\omega)$ over all sites in all systems of the ensemble.

Also from the local DOS we calculate the GIPR.
The usual inverse participation ratio is calculated from the amplitude of a single-particle wavefunction $\psi_{\alpha}$ at each site $i$ in the system:
$I_{\alpha} = \sum_i |\psi_{\alpha,i}|^4 / \left[ \sum_i |\psi_{\alpha,i}|^2 \right]^2$.
This is a standard measure of localization in non-interacting systems.
$I_{\alpha}$ is proportional to one over the number of sites on which $\psi_{\alpha}$ has nonzero weight.
$I_{\alpha}$ has a maximum value of one--indicating a state localized on a single site--and a minimum value of one over the system size--indicating an extended state.

However, in interacting systems single-particle states are not well defined.
The GIPR replaces the amplitude squared of a single-particle wavefunction at site $i$ with the local DOS at that site at a given energy:
\begin{eqnarray}
I(\omega) &=& \sum_i \rho_i^2(\omega) / \left[ \sum_i \rho_i(\omega)\right]^2
\label{ipr}
\end{eqnarray}
The GIPR quantifies the localization of a transition between many-body eitenstates.
It is proportional to one over the number of sites on which a single-particle transition from the ground state has nonzero weight.
In evaluating this, it is key that we know the many-body eigenstates themselves and not simply the Green's function,
because, when only the Green's function is known and only with finite energy resolution, the correspondence between the non-interacting limit of this expression and the usual inverse participation ratio is only well defined for zero disorder and infinite disorder but not for intermediate values.\cite{Murphy2011}
Here we may consider each transition individually, no matter how close they may be in energy.  
The local DOS at site $i$ is a sum of weighted $\delta$ functions: $\rho_{i}(\omega) = \sum_t w_{ti} \delta(\omega-E_t)$ where $E_t$ are the energies of single-particle transitions accessible from the ground state.  
By interpreting Eq.\ (\ref{ipr}) as $I(\omega) = \sum_i w_{ti}^2 /\left[ \sum_i w_{ti} \right]^2$ for $\omega=E_t$ and zero otherwise, the GIPR reduces to the usual IPR in the absence of interactions.

In constructing an ensemble-average of the GIPR as a function of energy, similar to our ensemble-average DOS,
a complication is that multiple transitions contribute to the DOS in the same energy window and the GIPR values for these transitions can vary widely.  
Our ensemble-average GIPR averages over all transitions $t$ in all systems $s$ in a given energy bin, weighting each GIPR value $I_{st}$ with the corresponding DOS contribution $\rho_{st}$:
\begin{eqnarray}
\langle I(\omega) \rangle &=& {\sum_s \sum_t I_{st}(\omega) \rho_{st}(\omega)\delta(\omega-\omega_{st})
\over \sum_s \sum_t \rho_{st} \delta(\omega-\omega_{st})}
\end{eqnarray}

\section{Results}
\label{sec-res}

We have examined the variation of the DOS and the GIPR 
both as a function of filling 
and as a function of system size.

Fig.\ \ref{filling} (a) shows the ensemble-average DOS and GIPR for ensembles of two-site systems at five different values of filling.  
The rough shape of the DOS in all cases is a broad, relatively flat distribution with a central, raised region. 
This structure is closely related to the atomic limit DOS in which the broad lower section comes from the contributions of singly-occupied sites and the central plateau comes from the contributions of empty and doubly-occupied sites.\cite{Wortis2011}
The presence of hopping introduces a sharp suppression at zero frequency known as a zero-bias anomaly.
The behavior of the zero-bias anomaly has been explored in some detail elsewhere.\cite{Chiesa2008,Wortis2010,Wortis2011}
As the system is doped below half filling, the central plateau moves to higher frequencies.
The zero-bias anomaly persists so long as the Fermi level falls inside the central plateau, but vanishes outside this range.

The focus of the present work is the coincidence of sharp peaks in the DOS at the edge of the central plateau with abrupt drops in the GIPR.  
This coincidence is highlighted in each panel of Fig.\ \ref{filling} (a) by vertical dotted lines.
% Add grey lines on the negative frequency side for the first three fillings?
Fig.\ \ref{filling} (b) helps to understand the coincidence of these two features in the simple case of no interactions.\cite{Johri2012b} 
Each point in the blue diamond represents a particular two-site system with specific values of $\epsilon_1$ and $\epsilon_2$.  
In the absence of interactions, each system contributes to the DOS at two energies, corresponding to the bonding and anti-bonding single-particle states of the system.  
The three hyperbolae on the graph show lines of constant energy of the anti-bonding state.
The DOS contribution (from anti-bonding states) at a given energy is equal to the length of the overlap of the corresponding hyperbola with the phase-space diamond.
At the energy labeled $E_{peak}$, the hyperbola is nearly tangent to the edges of the diamond.  Therefore the length of the overlap and hence the magnitude of the DOS is a maximum at this energy.
At slightly higher energies, the length of the overlap (and hence the magnitude of the DOS) drops sharply.  

This figure also demonstrates the connection with the GIPR.
Along a horizontal line through the center of the diamond, the potentials on the two sites are equal:  $\epsilon_1=\epsilon_2$.  
In the region around this line (shaded paler blue) the single-particle states have roughly equal weight on both sites and hence GIPR values near 0.5, the minimum possible value in a two-site system.
At $E_{peak}$ the hyperbola passes through systems with the full range of states, from maximally localized at the top and bottom to maximally extended in the center.  
The ensemble-average GIPR at this energy is therefore high.  In fact in non-interacting systems this is the inverse participation ratio maximum.
However for energies above $E_{peak}$ the hyperbola passes primarily through systems with maximally extended states.
Therefore, the peak in the DOS at $E_{peak}$ coincides with the lower boundary of a region with suppressed localization.

In interacting systems, the detailed picture is significantly more complicated.
The single-particle transitions available to the system depend on the number of particles in the ground state, and
therefore the diamond in Fig.\ \ref{filling} (b) must be divided into sections reflecting these different ground states.
For a given ground state there are multiple transitions each with a different constant energy curve.
A detailed discussion for the case of half-filling appears in Ref.\ [\onlinecite{Perera2015}].
Despite the added complexity, however, the basic picture remains the same:  Matches between the shapes of specific constant energy curves and the corresponding phase-space region result in peaks in the DOS, and moreover these peaks mark the edges of energy ranges with sharply suppressed GIPR.

Here we have examined the effect of doping away from half filling in ensembles of two-site systems.
In Fig.\ \ref{filling} (a) we see that the coincidence of DOS peaks and GIPR dips, marked by dotted lines, persists.
In particular, this is true on the high frequency side of the central plateau at all frequencies.
However, at filling 0.35 and below, where the Fermi level no longer falls inside the central plateau, neither a sharp feature in the DOS nor a deep drop in the GIPR appear.
What has occurred here is that the particular transition responsible for the peak is no longer present because the corresponding ground state no longer occurs.  
In particular, with sufficient under doping the 4-particle ground state and eventually the 3-particle ground state cease.  
The transition from the 3-particle ground state to the 2-particle triplet excited state plays the key role in the DOS peak on the left side of the central plateau.\cite{Perera2015}
When the 3-particle ground state is suppressed the peak disappears and with it the GIPR dip.

\begin{figure}
\includegraphics[width=\columnwidth]{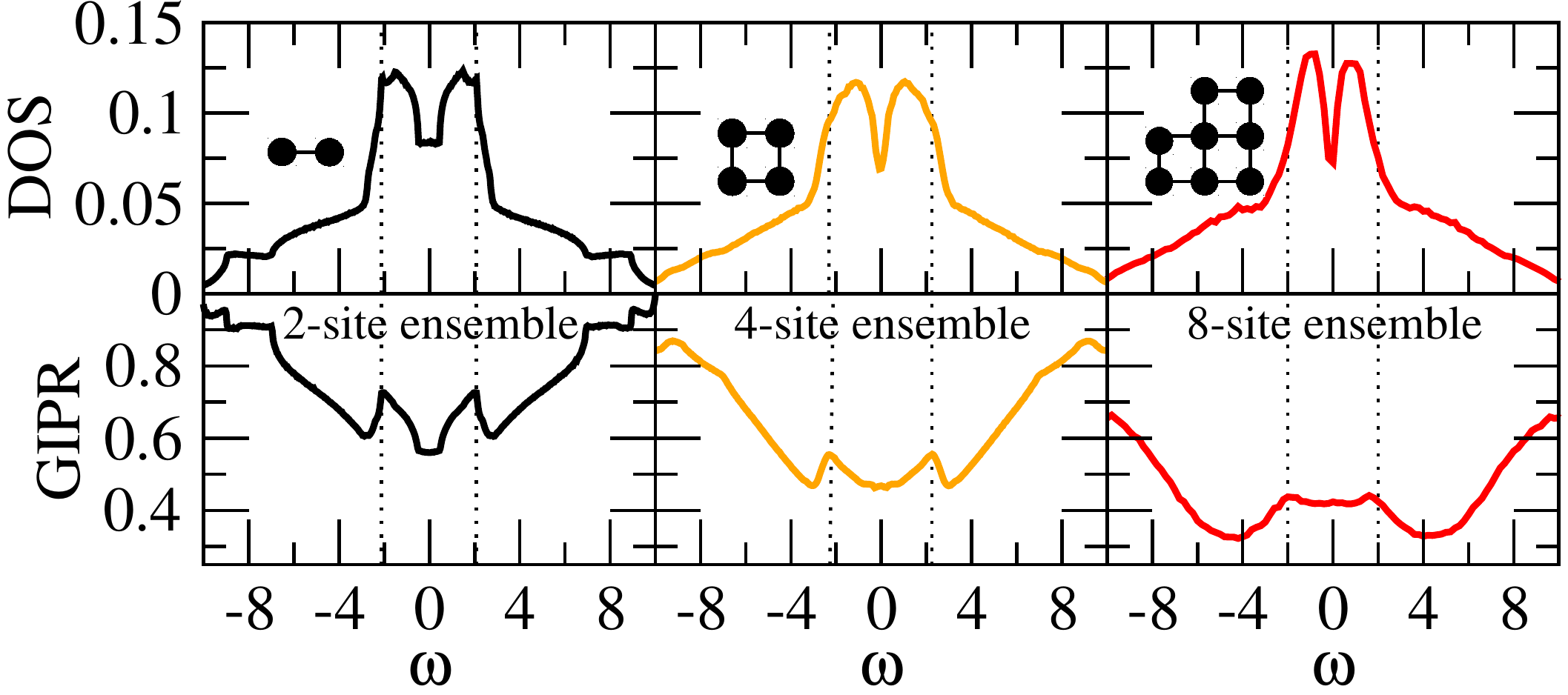} 
\caption{\label{size} (Color online) DOS and GIPR for ensembles of systems with two, four and eight sites.
All curves are at half filling with disorder strength $W/t=12$ and interaction strength $U/t=8$.
The two-site results average over an ensemble of ten million systems and have an energy resolution of $0.04t$.
The four-site results average over an ensemble of one million systems and have an energy resolution of $0.08t$.
The eight-site results average over an ensemble of 8300 systems and have an energy resolution of $0.12t$.
number of systems
}
\end{figure}

Fig.\ref{size} shows the ensemble-average DOS and GIPR at half filling in ensembles of systems of different sizes.
%For two-site and four-site systems, there is no distinction between a one-dimensional configuration and a two-dimensional one.  
%For the eight-site systems we use a two-dimensional Betts lattice as shown.
%However, for eight-site systems we considered both a one-dimensional chain with periodic boundary conditions and a two-dimensional Betts lattice also with periodic boundary conditions.
For systems larger than two sites, all sharp features in the DOS are lost with the notable exception of the zero-bias anomaly.  
For four-site systems, the analogue of the cartoon shown in Fig.\ \ref{filling} (b) begins with a four-dimensional diamond and for eight-site systems an eight-dimensional one.  
The lack of sharp features in the DOS suggests that the corresponding constant energy surfaces in these higher dimensional spaces no longer have surfaces which run nearly parallel with the boundaries of the phase space.
In the larger systems there are also a vastly increased number of transitions contributing, reducing the significance of individual transitions.

Despite the absence of peaks in the DOS in larger systems, the GIPR continues to show regions of depressed localization.
As in the two-site case, these appear just outside the central raised region in the DOS.
This greater robustness of the GIPR feature relative to the DOS peaks is similar to what was found in non-interacting systems.
Ref.\ [\onlinecite{Johri2012b}] argued that a singularity persists in the GIPR in the thermodynamic limit, whereas the fate of the DOS peak was less clear.
While exact calculations have severe size restrictions, there is evidence from alternative numerical techniques that the energy dependence of localization persists in even larger systems.
Ref.\ [\onlinecite{Andrade2009}] observed energy-dependent variation of localization in a study of the Anderson-Hubbard model using a single-site statistical dynamical mean-field theory approach\cite{Dobrosavljevic1997} with a slave-boson impurity solver which allows results on a $50\times50$ lattice.  % See Andrade Fig. 6
In particular, close to the Mott transition they observed that increasing interaction strength reduced localization at the Fermi level but {\em not} away from the Fermi level.  
This result is consistent with results in the two-site ensemble, suggesting that the energy-dependence of localization seen in small systems does indeed persist in large systems even if it is no longer flagged by peaks in the DOS.
% I'm not able to speak to whether the two-site results are consistent with Andrade away from the Fermi level.  Partly this is because I am not clear what they mean by $\omega/Z_{typ}=0.1$ and partly because once I had the specific frequency I would have to explicitly run the two-site ensemble at the two specific interaction strengths.  Nonetheless, the basic idea that you can get a drop in GIPR at $\omega=0$ but a rise away from zero is entirely consistent with what we see in the two-site ensemble.

The GIPR provides a length scale associated with a single-particle transition between the ground state of a many-body system and an excited state.  
The energy dependence we find would be reflected experimentally in a variation of the length scale over which the injection or removal of an electron would effect a disordered interacting system, with higher energy electrons connecting with longer length scales than lower energy electrons.

While the localization of single-particle transitions is distinct from the localization of many-body states, the two are connected.  
A transition between two many-body eigenstates can be expressed as the flipping of an Ising spin.\cite{Lychkovskiy2013}
When a system is many-body localized, it has been argued that these Ising spins may be chosen to be local.\cite{Huse2014} 
The range of GIPR values obtained provides some sense of the range of length scales associated with these locally conserved quantities, although the picture is incomplete because only transitions from the ground state are included.  

% I am not sure what to do with our 1D 8-site results.  Currently, I don't have sufficient confidence that they are correct to include them.

\section{Summary}
\label{sec-sum}

We have examined a connection, which has been seen in non-interacting systems and in ensembles of two-site interacting systems at half filling, between sharp peaks in the single-particle DOS and energy regions with suppressed localization.
We have shown that this connection persists away from half filling in ensembles of two-site systems,  
although it is possible to dope a system sufficiently that a key ground state no longer occurs, removing the transitions from this ground state and erasing both the DOS peak and the GIPR dip simultaneously.
The magnitude of the doping needed for this to happen depends on the magnitudes of the interaction and disorder strengths.  
We have also looked at the dependence of the effect on the size of the systems.  
We find that, for ensembles of systems larger than two sites, sharp peaks in the DOS no longer appear.  This is in contrast to the non-interacting case.\cite{Johri2012b}
As for the GIPR, we find that despite the lack of DOS features a region of suppressed GIPR does persist up to ensembles of eight-site systems.
% Mention Andrade again.

%Extra note:  After reading Perera paper, Bill suggested (15-05-18 email) exploring whether the cusps can be seen in a DMFT + CPA calculation.  Why?  We know CPA doesn't capture localization, but we also know that CPA usually gets the DOS right.  Therefore, if the DMFT+CPA gets the DOS peaks, they are not related to localization.  If it doesn't, that's indirect support for there being a connection.  As for the GIPR, something similar holds, but I don't have any opinions on how well CPA should do for the GIPR.  Note that the place to start with this is to do the non-interacting case first.  All the same arguments hold, and the calculation is a lot easier.  One of Bill's concerns is that we don't really know what the connection is between the GIPR and localization.  That is, I think we are confident there's a connection in non-interacting systems, but for interacting systems it's not clear.

\acknowledgements
We gratefully acknowledge coding advice from W.A.\ Atkinson and support by the Natural Sciences and Engineering Research Council (NSERC) of Canada.  This work was made possible by the facilities of the Shared Hierarchical Academic Research Computing Network (SHARCNET).

%\bibliography{Daley}

\end{document}